\documentclass{article}
\usepackage{microtype}
\usepackage{graphicx}
\usepackage{subcaption}
\usepackage{booktabs} 

\usepackage{hyperref}
\usepackage{float}
\usepackage{graphicx}  

\usepackage[preprint]{icml2026}

\usepackage{amsmath}
\usepackage{amssymb}
\usepackage{mathtools}
\usepackage{amsthm}
\usepackage{pifont}    
\usepackage{xcolor}    
\usepackage{longtable}
\usepackage{enumitem}
\usepackage{listings}
\usepackage{xcolor}
\usepackage{cuted}
\usepackage{times}
\usepackage{latexsym}
\usepackage{booktabs}
\usepackage{tabularx}
\usepackage{threeparttable}
\usepackage{fancyvrb}
\usepackage{tcolorbox}
\usepackage{longtable}
\usepackage{array}
\usepackage{fvextra} 
\tcbuselibrary{breakable}
\usepackage{lipsum} 
\usepackage[T1]{fontenc}

\usepackage[utf8]{inputenc}
\usepackage{booktabs}

\usepackage{microtype}

\usepackage{inconsolata}

\usepackage[capitalize,noabbrev]{cleveref}

\newcolumntype{L}[1]{>{\raggedright\arraybackslash}p{#1}}
\newcolumntype{C}[1]{>{\centering\arraybackslash}p{#1}}

\theoremstyle{plain}

\theoremstyle{definition}

\theoremstyle{remark}

\usepackage[textsize=tiny]{todonotes}

\icmltitlerunning{CaReCoS: A Spectrogram based Visual Benchmark for Cardiac, Respiratory and Cough Sounds}

\begin{document}
\twocolumn[
  \icmltitle{CaReCoS: A Spectrogram based Visual Benchmark for Cardiac, Respiratory and Cough Sounds}

  \icmlsetsymbol{equal}{*}

  \begin{icmlauthorlist}
    \icmlauthor{Harshit Rajgarhia}{centific}
    \icmlauthor{Shuubham Ojha}{umd,centific}
    \icmlauthor{Asif Shaik}{centific}
    \icmlauthor{Akhil Pothanapalli}{centific}
    \icmlauthor{Rachuri Lokesh}{centific}
    \icmlauthor{Abhishek Mukherji}{centific}
    \icmlauthor{Prasanna Desikan}{centific}
  \end{icmlauthorlist}

  \icmlaffiliation{centific}{Centific Global Solutions Inc.}
  \icmlaffiliation{umd}{University of Maryland, College Park, MD, USA}

  \icmlcorrespondingauthor{Harshit Rajgarhia}{harshit.rajgarhia@centific.com}

  \icmlkeywords{Machine Learning, ICML}

  \vskip 0.3in
]

\printAffiliationsAndNotice{}  

\begin{abstract}

Medical acoustic signals such as respiratory sounds, cardiac auscultations, and cough audio carry rich diagnostic information, yet no existing benchmark evaluates multimodal reasoning over their spectrogram representations. We address both gaps with \textbf{CaReCoS}, a benchmark pairing clinically grounded questions with mel-spectrogram images derived from seven medical audio datasets. Evaluating 9 state-of-the-art vision and omni models, we find that all struggle with fine-grained acoustic features encoded in spectrograms: no model reliably combines visual pattern recognition with medical knowledge, achieving a maximum accuracy of 51.2\%, underscoring the need for training on medical sound visualizations.

\end{abstract}

\section{Introduction}
\label{sec:intro}

Body-produced acoustic signals, including respiratory sounds such as wheezes, crackles, and rhonchi, and cardiac murmurs of varying timing, pitch, and intensity, have long served as a fundamental diagnostic modality \cite{rocha2019icbhi, oliveira2022circor}. Despite their widespread clinical use, automated analysis of these signals remains constrained by small datasets, noisy recordings, and poor cross-setting generalization.
 
Large multimodal foundation models offer an opportunity to change this. Audio LLMs such as SALMONN \cite{tang2023salmonn} and Qwen2-Audio \cite{chu2024qwen2audio}, vision-language models (VLMs) such as GPT-4V \cite{openai2023gpt4} and LLaVA \cite{liu2023llava}, and omni models such as GPT-4o \cite{openai2024gpt4o} and Qwen2.5-Omni \cite{qwen2025omni} can process waveforms, images, or both yet whether any can genuinely reason about medical acoustic content remains unexplored.
 
A central obstacle is the absence of structured benchmarks. Audio benchmarks such as AudioBench \cite{wang2024audiobench} and MMAU \cite{sakshi2024mmau} contain no medical content; medical datasets such as ICBHI 2017 \cite{rocha2019icbhi} and CirCor \cite{oliveira2022circor} provide labels but no reasoning-oriented QA pairs. Finally, no benchmark evaluates models through spectrograms, the de-facto visual representation of sound in clinical practice. We close these gaps with two contributions:
 
\begin{enumerate}
    \item \textbf{CaReCoS}, a two-tiered benchmark comprising \emph{explicit} questions whose answers are anchored to dataset metadata and \emph{inferred} questions that demand multi-step clinical reasoning, with ground-truth answers generated and verified by Gemini 3 Flash.
 
    \item \textbf{A spectrogram-based evaluation protocol} in which all 9 evaluated models receive a mel-spectrogram image of each audio clip alongside the question text, enabling genuine cross-modal understanding.
    
\end{enumerate}

\begin{figure*}
\centering
    \includegraphics[width=0.8\textwidth]{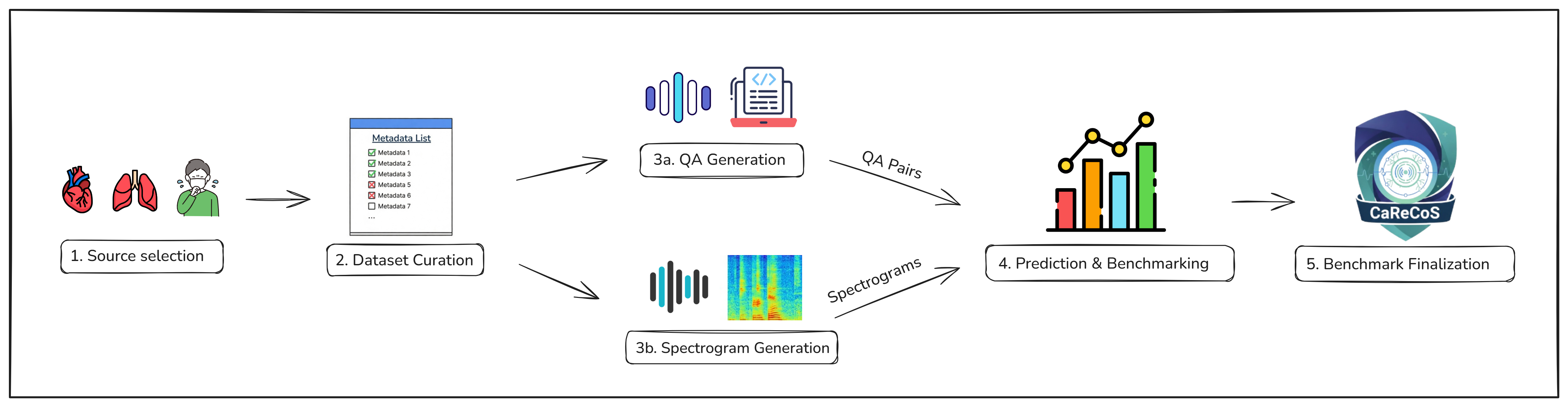}
    \caption{A flow diagram of the proposed pipeline for benchmark generation.}
\end{figure*}

\section{Related Work}
\label{sec:related}

\subsection{General Audio Reasoning: Models and Benchmarks}
\label{subsec:audio}

Audio-conditioned large language models (LALMs) have advanced rapidly, from SALMONN \cite{tang2023salmonn}, which fuses speech and general audio encoders for ASR, captioning, and zero-shot QA, to the Qwen-Audio family \cite{chu2023qwenaudio, chu2024qwen2audio} scaling to 7B parameters with state-of-the-art instruction-following.
Evaluation benchmarks have grown in parallel: AIR-Bench \cite{yang2024airbench} covers generative comprehension across speech, sound, and music; AudioBench \cite{wang2024audiobench} adds model-as-judge scoring; and MMAU \cite{sakshi2024mmau} poses 10,000 expert-level questions yet even Gemini 1.5 Pro reaches only ~66\% accuracy. Omni-modal architectures such as GPT-4o \cite{openai2024gpt4o}, Qwen2.5-Omni \cite{qwen2025omni}, and Gemini \cite{team2024gemini} unify audio, vision, and text, but none have been evaluated on medical acoustics. Critically, no existing benchmark uses spectrogram images as a primary input, despite spectrograms being the standard visualization in clinical decision-support software \cite{fraiwan2021dataset, kandasamy2011monitoring}.

\subsection{General Image Reasoning: Models and Benchmarks}
\label{subsec:vision}

Vision-language models (VLMs) have scaled rapidly through instruction-tuning on web-scale image-text data, from LLaVA \cite{liu2023llava} through GPT-4V \cite{openai2023gpt4} and Gemini \cite{team2024gemini} to open-source alternatives such as Qwen2.5-VL \cite{qwen2025vl} and InternVL2.
Benchmarks have grown correspondingly harder: MMBench \cite{liu2023mmbench} assesses 20 capability dimensions; MMMU \cite{yue2024mmmu} targets expert-level multimodal reasoning across six disciplines (GPT-4V ~56\%); MME \cite{fu2023mme} spans 14 perception and cognition subtasks; and MMMU-Pro \cite{yue2024mmmupro} and NaturalBench \cite{li2024naturalbench} reveal that many gains reflect language bias rather than genuine visual understanding. Critically, these benchmarks draw almost exclusively from photographs, charts, and diagrams. Scientific visualizations such as spectrograms, ECGs, and spirometry traces are virtually absent from both training and evaluation corpora.

\subsection{Medical Sound Reasoning}
\label{subsec:medical}
\begin{figure*}
\centering
    \includegraphics[width=0.8\textwidth]{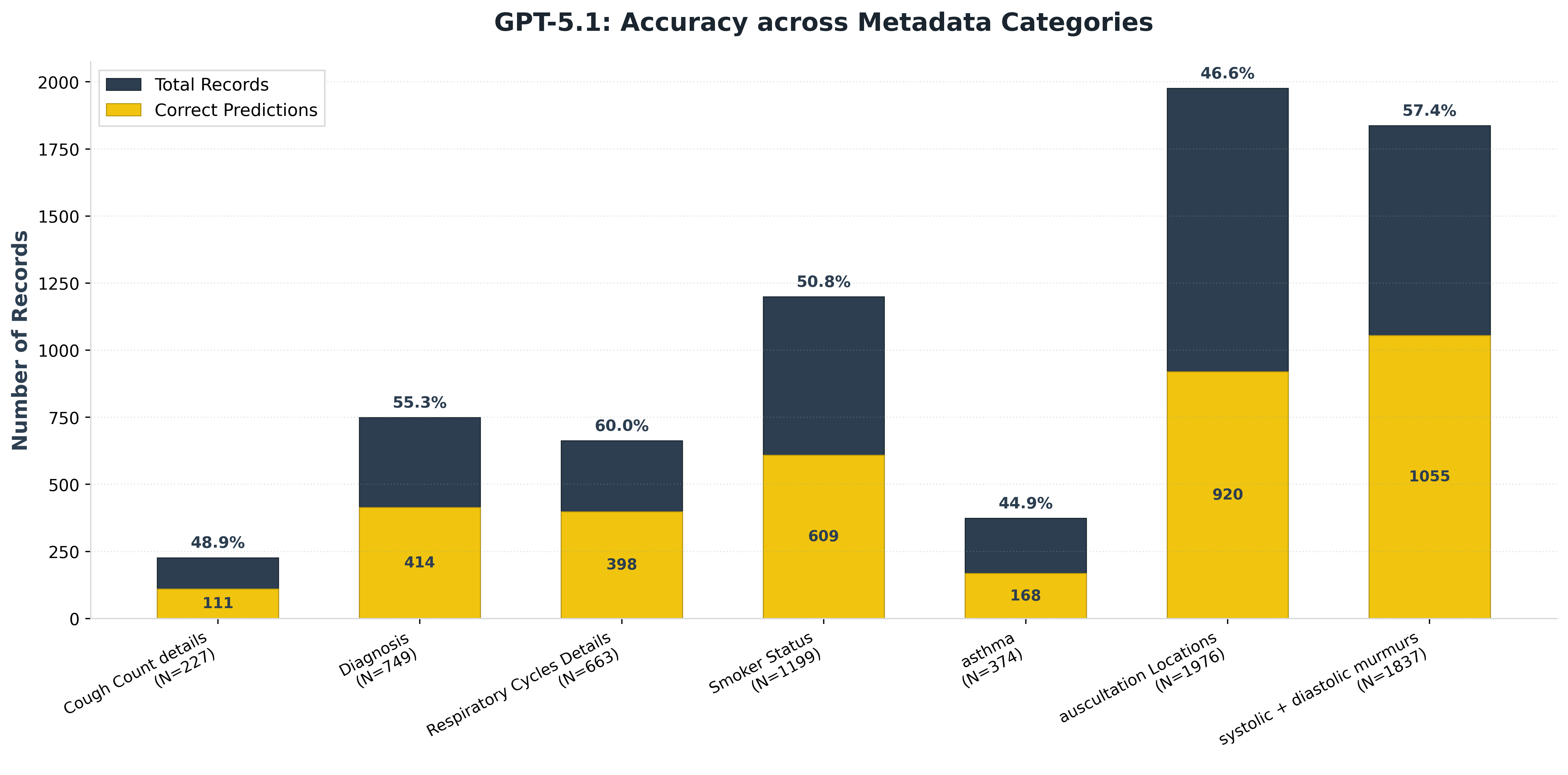}
    \caption{Number of total QA pairs in CaReCoS and relative accuracy of GPT 5.1}
\end{figure*}

Traditional respiratory and cardiac sound analysis relies on hand-crafted features (MFCCs, wavelet coefficients, log-mel spectrograms) fed into CNN or RNN classifiers. Datasets such as ICBHI 2017 \cite{rocha2019icbhi} and PhysioNet CirCor \cite{oliveira2022circor} have driven progress but frame the task as classification, not reasoning.
Foundation model approaches are emerging: HeAR \cite{baur2024hear}, a self-supervised encoder pre-trained on 313M health acoustic clips, and OPERA \cite{zhang2024opera} provide pre-trained representations but lack reasoning capability. RespLLM \cite{zhang2024respllm} moves toward instruction-following by fusing respiratory audio with demographic text, showing zero-shot generalisation to unseen conditions, while fine-tuning Qwen2-Audio on phonocardiogram features \cite{anonymous2025pcg} approaches supervised baselines.
Three gaps remain, which our work addresses: no benchmark exists for medical sound \emph{reasoning} (as opposed to classification); no work has evaluated general-purpose foundation models on medical sounds zero-shot; and the spectrogram as a visual interface has not been benchmarked despite its routine clinical use. 


\begin{table*}[ht]
\centering
\caption{Summary of benchamrking resuts for various models. The highest accuracies in every row is bolded and the highest accuracies in every column is underlined.}
\small
\begin{tabular}{l c c c c c c c c}
\toprule
\textbf{Model} & \textbf{CirCor} & \textbf{Coswara} & \textbf{ICBHI} & \textbf{KAUH} & \textbf{SPRSound} & \textbf{ZCH} & \textbf{Cough Count} & \textbf{Weighted Avg} \\
 & (1600) & (1512) & (672) & (336) & (672) & (80) & (168) & \\
\midrule
Gemini 2.5 Flash  & 20.9 & 22.1 & 32.0 & 23.8 & 29.9 & 27.5 & \textbf{38.1} & 24.8 \\
Gemini 2.5 Pro  & \textbf{48.0} & 42.9 & 39.3 & 41.7 & 39.7 & 28.7 & 38.7 & 43.2 \\
GPT-5.1  & \underline{49.8} & \underline{47.6} & \underline{\textbf{51.2}} & \underline{47.9} & \underline{49.9} & \underline{31.2} & \underline{46.4} & \underline{48.8} \\
Kimi-VL  & 26.6 & \textbf{27.4} & 24.7 & 17.0 & 22.0 & 16.2 & 21.4 & 25.6 \\
LLaVA-OneVision  & 8.7 & \textbf{21.5} & 15.5 & 8.3 & 12.6 & 15.0 & 8.9 & 14.1 \\
Med-LLaVA   & 4.2 & 6.2 & \textbf{14.1} & 4.4 & 10.9 & 8.8 & 6.5 & 7.1 \\
MedGemma  & 18.6 & \textbf{25.9} & 16.4 & 17.3 & 18.1 & 13.8 & 8.4 & 19.9\\
Qwen-Omni 7B  & 12.4 & \textbf{26.6} & 20.8 & 11.0 & 17.1 & 15.0 & 13.1 & 18.4 \\
Qwen-VL 2.5 & 5.4 & 7.6 & \textbf{9.6} & 2.7 & 7.6 & 2.6 & 4.3 & 6.6 \\
\bottomrule
\end{tabular}
\end{table*}

\section{Dataset Design}
\label{sec:dataset_design}

Our benchmark draws on seven publicly available datasets spanning three medical auscultation sound types: lung sounds, heart sounds, and cough sounds.
For lung sounds, we use three datasets: ICBHI \cite{rocha2019open}, which provides respiratory cycle recordings annotated with chest location, crackle/wheeze presence, and patient diagnosis; KAUH \cite{fraiwan2021dataset}, containing recordings with sound type, diagnosis, and recording location annotations; and SPRSound \cite{zhang2022sprsound}, which includes both record-level and event-level annotations (e.g., rhonchi, stridor, fine crackle) with recording location. For heart sounds, we use CirCor \cite{oliveira2022circor}, providing phonocardiogram recordings with detailed murmur characterization (timing, shape, grading, pitch, quality) across four standard auscultation sites, and ZCH \cite{zhang2024zchsound}, containing neonatal and pediatric recordings with structural diagnoses (e.g., Atrial Septal Defect (ASD), Ventricular Septal Defect (VSD), Patent Ductus Arteriosus (PDA)). For cough sounds, we use Coswara \cite{sharma2020coswara}, providing cough recordings with clinical metadata including asthma, COVID-19, and smoking status, and the EPFL Multimodal Cough Detection dataset \cite{orlandic2023multimodal}, which provides cough event timestamps from which we derive temporal features such as cough count, per-cough duration, inter-cough intervals etc. For each recording, Gemini 3 Flash was provided the metadata and waveform to generate four QA pairs: two explicit (direct retrieval of annotated attributes) and two inferred (clinical reasoning grounded in the spectrogram). 
Since, explicit ground truths are drawn directly from dataset metadata they require no additional human validation. For inferred questions, human evaluation would demand clinicians skilled in both auscultation and spectrogram interpretation, a combination difficult to source at scale.





\section {Experiments}
\label{ref:Experiments}

Our evaluation pipeline consists of two stages: model prediction and automated judgment. In the prediction stage, each audio recording is converted to a mel spectrogram (nfft=1024, hop length=512, n mels=256), which serves as visual input to the model alongside a generated question. The model is prompted to infer the sample rate from the spectrogram's y-axis and respond in one to two sentences. This formulation enables benchmarking vision-language models on medical audio understanding without requiring native audio input support.
In the judgment stage, we employ an LLM-as-a-judge framework with category-specific evaluation criteria. For explicit questions, the judge applies strict factual matching: the model response must convey the same factual information as the reference answer with no contradictions. Credit is withheld if the response introduces facts not present in the ground truth or if key clinical values (e.g., auscultation location, murmur grade, diagnosis) are incorrect. Minor surface-level rephrasing is permitted provided the core medical facts are preserved. For inferred questions, the judge applies relaxed semantic evaluation: credit is awarded if the response conveys the same core clinical concept, mechanism, or conclusion, even if phrased differently. Minor omissions are accepted as long as the primary medical reasoning is correct, while clinically significant errors or vague responses (inability to interpret the spectrogram) receive no credit.

GPT-5.1 achieves the highest weighted average accuracy at 48.8\%, leading across most datasets including CirCor (49.8\%), ICBHI (51.2\%), and cough count (46.4\%), making it the strongest performer on spectrogram-based medical QA overall. Gemini 2.5 Pro follows as the second-best model with a 43.2\% weighted average, showing particular strength on the CirCor heart sound dataset (48.0\%) but dropping notably on ZCH (28.7\%). Gemini 2.5 Flash (24.8\%) and Kimi-VL (25.6\%) occupy a middle tier, with Flash showing relative strength on ICBHI (32.0\%) and cough count (38.1\%) lung-related tasks. Among the medical domain-specific models, MedGemma (19.9\%) outperforms Med-LLaVA (7.1\%), though both trail significantly behind the general-purpose frontier models, suggesting that current medical vision-language fine-tuning does not transfer well to auscultation spectrogram interpretation. Qwen-VL-2.5 records the lowest overall accuracy at 6.6\%, and open-source models generally (LLaVA-OneVision at 14.1\%, Qwen-Omni-7B at 18.4\%) lag substantially behind proprietary systems. Notably, all models perform worst on the ZCH neonatal heart sound dataset, indicating that pediatric cardiac spectrogram analysis remains particularly challenging. 

As an ablation, we also evaluated three omni models (Qwen2.5-Omni-7B, Gemini 2.5 Flash, Gemini 2.5 Pro) using raw audio instead of spectrograms. Gemini 2.5 Pro achieves 42.0\% overall with audio, comparable to its 43.2\% spectrogram accuracy, while Gemini 2.5 Flash jumps from 24.8\% to 40.7\%, a gain driven primarily by a 42\% spectrogram refusal rate that largely disappears with audio input. Qwen2.5-Omni-7B shows negligible change (18.8\% vs 18.4\%). 


\section{Discussion}
\label{ref:Discussion}
  
The limited performance of medically finetuned models stems from a domain mismatch: the vision encoders in MedGemma (MedSigLIP) \cite{sellergren2025medgemma} and LLaVA-Med (CLIP) \cite{radford2021clip} have learned to detect anatomical edges, tissue textures, and spatial organ relationships, features absent from spectrograms. Clinically meaningful spectrogram patterns such as crescendo-decrescendo murmur envelopes have no visual analogue in the radiographs and pathology slides that dominate medical vision training sets. 
 
Table~\ref{tab:table2} shows the contrasting performance of GPT 5.1 on inferred versus explicit questions. Inspection of 100 questions per category reveals systematic errors on explicit questions: the model underestimates murmur duration (labeling holosystolic as mid-systolic), overestimates intensity grading (predicting grade II--III for grade I/VI), and confuses quality descriptors (harsh for blowing). 
 
For inferred questions, the model defaults to the most common textbook association, missing alternative diagnoses: given a harsh holosystolic murmur at the tricuspid area it identifies tricuspid regurgitation but overlooks ventricular septal defect, and shows a left-sided bias, attributing low-pitched blowing murmurs to mitral regurgitation even when context suggests right-sided pathology. These complementary weaknesse, perceptual feature extraction in the explicit setting and multi-feature clinical integration in the inferred setting indicate that broad cardiological knowledge alone is insufficient without the ability to weigh multiple acoustic features jointly.
 
\begin{figure}
\centering
    \includegraphics[width=0.5\textwidth]{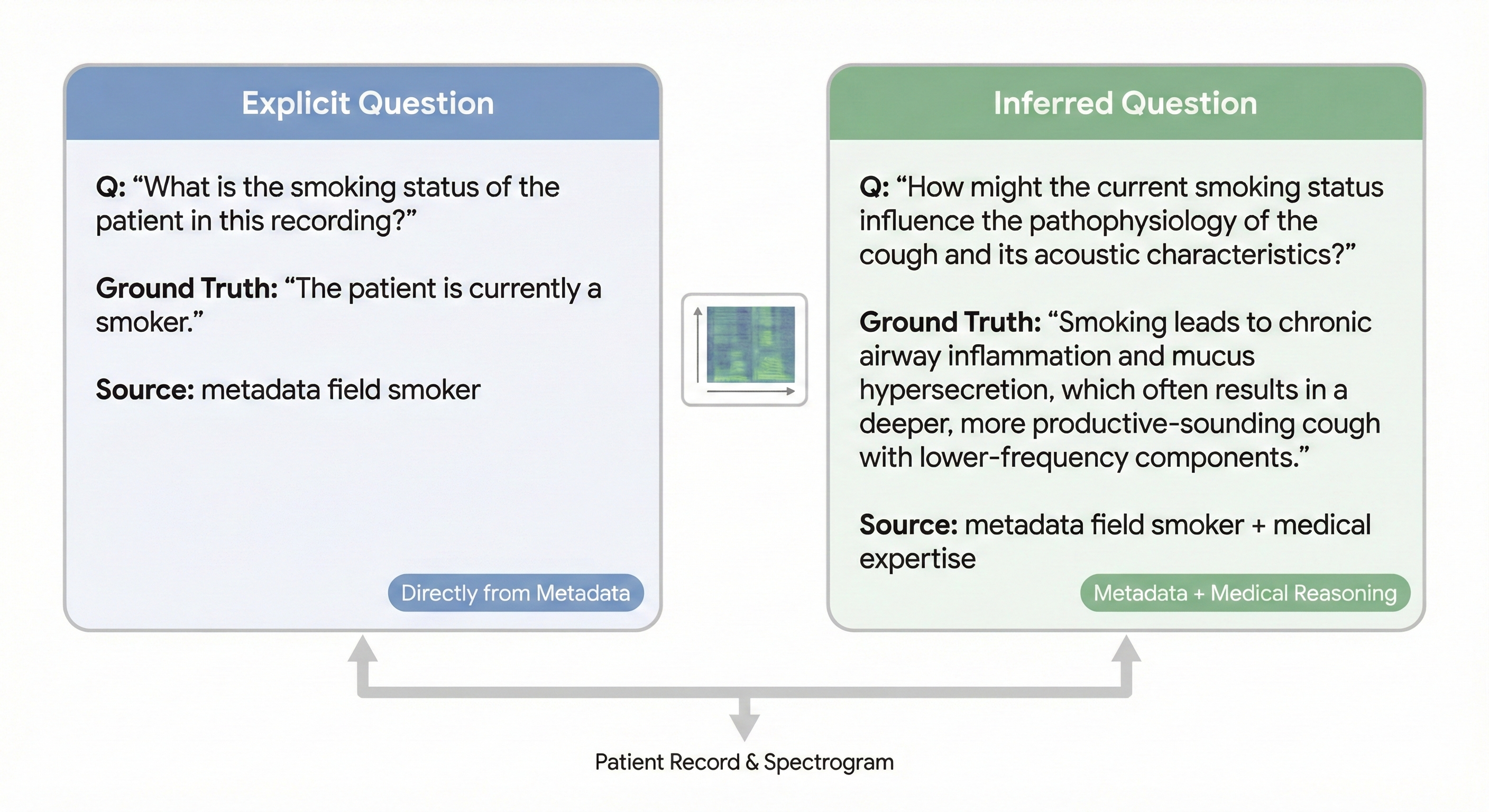}
    \caption{Example inferred and explicit questions.} 
\end{figure}
 
Qualitative error analysis validates our LLM-as-a-judge framework over embedding-based metrics. Models produce clinically fluent but factually incorrect responses, e.g., describing a murmur as ``crescendo-decrescendo, grade III/VI'' when the ground truth is ``plateau shape, grade II/VI'' that achieve BERT similarity scores above 0.95 despite fundamental clinical errors. Cross-domain confusion compounds this: a model labeling a COPD lung recording as ``normal heart sound'' at the ``left lower sternal border'' scores 0.92 against the respiratory ground truth due to shared clinical vocabulary. Conversely, honest uncertainty statements (``cannot be determined from the spectrogram alone'') score only 0.90, penalizing candor relative to confident hallucination. These failure modes confirm that clinical QA demands exact factual matching that surface-level similarity cannot provide.

\begin{table}[ht]
\centering
\caption{Accuracy comparison of inferred vs explicit questions. 2520 questions in each category.}
\label{tab:table2}
\begin{tabular}{lccc}
\hline
\textbf{Model} & \textbf{Inferred} & \textbf{Explicit} & \textbf{Weighted Avg} \\
\hline
Gemini 2.5 Flash   & 42.5  &  7.1   & 24.8 \\
Gemini 2.5 Pro     & 71.2  & \textbf{15.1}  & 43.2 \\
GPT 5.1              & \textbf{84.6}  & 13.0 & \textbf{48.8} \\
Kimi VL             & 36.9  & 13.1   & 25.0 \\
LLaVA One Vision   & 18.7   &  9.4   & 14.0 \\
Med LLaVA           &  8.6   &  5.7  &  6.8 \\
MedGemma             & 30.2   &  9.6  & 19.9 \\
Qwen Omni 7B       & 25.2  & 11.5   & 18.4 \\
Qwen VL 2.5         & 13.5   &  0.6   &  5.8 \\
\hline
\end{tabular}
\end{table}

\section{Conclusions}

We introduced CaReCoS, a spectrogram-based visual benchmark for medical sound reasoning spanning cardiac, respiratory, and cough domains, pairing mel spectrograms from seven clinical datasets with clinically grounded QA pairs. GPT-5.1, achieves only 48.8\% weighted accuracy. MedGemma and Med-LLaVA underperform general-purpose counterparts highlighting a domain mismatch between anatomical medical imagery and time-frequency representations. The gap between inferred and explicit accuracy shows that models struggle to extract precise acoustic features from spectrograms despite broad clinical knowledge. 


\label{ref:Conslusion}

\bibliography{mybib}

@inproceedings{tang2023salmonn,
  title        = {{SALMONN}: Towards Generic Hearing Abilities for Large Language Models},
  author       = {Tang, Changli and Yu, Wenyi and Sun, Guangzhi and Chen, Xianzhao and
                  Tan, Tian and Li, Wei and Lu, Lu and Ma, Zejun and Zhang, Chao},
  booktitle    = {The Twelfth International Conference on Learning Representations (ICLR)},
  year         = {2024},
  url          = {https://arxiv.org/abs/2310.13289}
}

@article{chu2023qwenaudio,
  title        = {Qwen-Audio: Advancing Universal Audio Understanding via Unified
                  Large-Scale Audio-Language Models},
  author       = {Chu, Yunfei and Xu, Jin and Zhou, Xiaohuan and Yang, Qian and
                  Zhang, Shiliang and Yan, Zhijie and Zhou, Chang and Zhou, Jingren},
  journal      = {arXiv preprint arXiv:2311.07919},
  year         = {2023}
}

@article{chu2024qwen2audio,
  title        = {Qwen2-Audio Technical Report},
  author       = {Chu, Yunfei and Xu, Jin and Yang, Qian and Wei, Haojie and
                  Wei, Xipin and Guo, Zhifang and Leng, Yichong and Lv, Yuanjun and
                  He, Jinzheng and Lin, Junyang and Zhou, Chang and Zhou, Jingren},
  journal      = {arXiv preprint arXiv:2407.10759},
  year         = {2024}
}

@article{qwen2025omni,
  title        = {Qwen2.5-Omni Technical Report},
  author       = {{Qwen Team}},
  journal      = {arXiv preprint arXiv:2503.20215},
  year         = {2025}
}

@inproceedings{wang2024audiobench,
  title        = {{AudioBench}: A Universal Benchmark for Audio Large Language Models},
  author       = {Wang, Bin and Zou, Xunlong and Lin, Geyu and Sun, Shuo and
                  Liu, Zhuohan and Zhang, Wenyu and Liu, Zhengyuan and
                  Aw, AiTi and Chen, Nancy F.},
  booktitle    = {Proceedings of the 2025 Annual Conference of the North American
                  Chapter of the Association for Computational Linguistics (NAACL)},
  year         = {2025},
  url          = {https://arxiv.org/abs/2406.16020}
}

@inproceedings{sakshi2024mmau,
  title        = {{MMAU}: A Massive Multi-Task Audio Understanding and Reasoning Benchmark},
  author       = {Sakshi, S. and Tyagi, Utkarsh and Kumar, Sonal and Seth, Ashish and
                  Selvakumar, Ramaneswaran and Nieto, Oriol and Duraiswami, Ramani and
                  Ghosh, Sreyan and Manocha, Dinesh},
  booktitle    = {The Thirteenth International Conference on Learning Representations (ICLR)},
  year         = {2025},
  url          = {https://arxiv.org/abs/2410.19168}
}

@article{yang2024airbench,
  title        = {{AIR-Bench}: Benchmarking Large Audio-Language Models via
                  Generative Comprehension},
  author       = {Yang, Qian and Xu, Jin and Liu, Wenrui and Chu, Yunfei and
                  Jiang, Ziyue and Zhou, Xiaohuan and Leng, Yichong and Lv, Yuanjun and
                  Zhao, Zhou and Zhou, Chang and others},
  journal      = {arXiv preprint arXiv:2402.07729},
  year         = {2024}
}

@article{openai2023gpt4,
  title        = {{GPT-4} Technical Report},
  author       = {{OpenAI}},
  journal      = {arXiv preprint arXiv:2303.08774},
  year         = {2023}
}

@article{openai2024gpt4o,
  title        = {{GPT-4o} System Card},
  author       = {{OpenAI}},
  journal      = {arXiv preprint arXiv:2410.21276},
  year         = {2024}
}

@inproceedings{liu2023llava,
  title        = {Visual Instruction Tuning},
  author       = {Liu, Haotian and Li, Chunyuan and Wu, Qingyang and Lee, Yong Jae},
  booktitle    = {Advances in Neural Information Processing Systems (NeurIPS)},
  volume       = {36},
  year         = {2023}
}

@article{qwen2025vl,
  title        = {Qwen2.5-{VL} Technical Report},
  author       = {{Qwen Team}},
  journal      = {arXiv preprint arXiv:2502.13923},
  year         = {2025}
}

@article{team2024gemini,
  title        = {Gemini 2.5: Pushing the Frontier with Advanced Reasoning,
                  Multimodality, Long Context, and Next Generation Agentic Capabilities},
  author       = {{Gemini Team, Google}},
  journal      = {arXiv preprint arXiv:2507.06261},
  year         = {2025}
}

@inproceedings{yue2024mmmu,
  title        = {{MMMU}: A Massive Multi-discipline Multimodal Understanding and
                  Reasoning Benchmark for Expert {AGI}},
  author       = {Yue, Xiang and Ni, Yuansheng and Zhang, Kai and Zheng, Tianyu and
                  Liu, Ruoqi and Zhang, Ge and Stevens, Samuel and Jiang, Dongfu and
                  Ren, Weiming and Sun, Yuxuan and others},
  booktitle    = {Proceedings of the IEEE/CVF Conference on Computer Vision and
                  Pattern Recognition (CVPR)},
  year         = {2024}
}

@article{yue2024mmmupro,
  title        = {{MMMU-Pro}: A More Robust Multi-discipline Multimodal Understanding
                  Benchmark},
  author       = {Yue, Xiang and others},
  journal      = {arXiv preprint arXiv:2409.02813},
  year         = {2024}
}

@inproceedings{liu2023mmbench,
  title        = {{MMBench}: Is Your Multi-modal Model an All-around Player?},
  author       = {Liu, Yuan and Duan, Haodong and Zhang, Yuanhan and Li, Bo and
                  Zhang, Songyang and Zhao, Wangbo and Yuan, Yike and Wang, Jiaqi and
                  He, Conghui and Liu, Ziwei and Chen, Kai and Lin, Dahua},
  booktitle    = {European Conference on Computer Vision (ECCV)},
  year         = {2024}
}

@inproceedings{fu2023mme,
  title        = {{MME}: A Comprehensive Evaluation Benchmark for Multimodal Large
                  Language Models},
  author       = {Fu, Chaoyou and Chen, Peixian and Shen, Yunhang and Qin, Yulei and
                  Zhang, Mengdan and Lin, Xu and Yang, Jinrui and Zheng, Xiawu and
                  Li, Ke and Sun, Xing and others},
  journal      = {arXiv preprint arXiv:2306.13394},
  year         = {2023}
}

@inproceedings{li2024naturalbench,
  title        = {{NaturalBench}: Evaluating Vision-Language Models on Natural
                  Adversarial Samples},
  author       = {Li, Baiqi and others},
  booktitle    = {Advances in Neural Information Processing Systems (NeurIPS),
                  Datasets and Benchmarks Track},
  year         = {2024}
}

@article{rocha2019icbhi,
  title        = {An Open Access Database for the Evaluation of Respiratory Sound
                  Classification Algorithms},
  author       = {Rocha, Bruno M. and Filos, Dimitris and Mendes, Lu{\'\i}s and
                  Serbes, Gorkem and Ulukaya, Sezer and Kahya, Yasemin P. and
                  Jakovljevic, Nikola and Turukalo, Tatjana L. and Vogiatzis, Ioannis M. and
                  Perantoni, Eleni and others},
  journal      = {Physiological Measurement},
  volume       = {40},
  number       = {3},
  pages        = {035001},
  year         = {2019},
  publisher    = {IOP Publishing}
}

@article{rocha2019open,
  title        = {An Open Access Database for the Evaluation of Respiratory Sound
                  Classification Algorithms},
  author       = {Rocha, Bruno M. and Filos, Dimitris and Mendes, Lu{\'\i}s and
                  Serbes, Gorkem and Ulukaya, Sezer and Kahya, Yasemin P. and
                  Jakovljevic, Nikola and Turukalo, Tatjana L. and Vogiatzis, Ioannis M. and
                  Perantoni, Eleni and others},
  journal      = {Physiological Measurement},
  volume       = {40},
  number       = {3},
  pages        = {035001},
  year         = {2019},
  publisher    = {IOP Publishing}
}

@article{oliveira2022circor,
  title        = {The {CirCor DigiScope} Dataset: From Murmur Detection to Murmur
                  Classification},
  author       = {Oliveira, Jorge and Renna, Francesco and Costa, Paulo Dias and
                  Nogueira, Marcelo and Oliveira, Cristina and Ferreira, Carlos and
                  Jorge, Al{\'\i}pio and Mattos, Sandra and Hatem, Thamine and
                  Tavares, Thiago and Elola, Andoni and Rad, Ali Bahrami and
                  Sameni, Reza and Clifford, Gari D. and Coimbra, Miguel T.},
  journal      = {IEEE Journal of Biomedical and Health Informatics},
  volume       = {26},
  number       = {6},
  pages        = {2524--2535},
  year         = {2022},
  publisher    = {IEEE},
  doi          = {10.1109/JBHI.2021.3137048}
}

@article{zhang2024zchsound,
  title        = {{ZCHSound}: Open-source {ZJU} Paediatric Heart Sound Database
                  with Congenital Heart Disease},
  author       = {Zhang, Yuhang and Zhang, Qing and Zhang, Jing and Yuan, Jiajun and
                  Huang, Huajie and Zhang, Baoqin and Lv, Gaomei and Lin, Shuzhu and
                  Wang, Na and Liu, Xin and others},
  journal      = {IEEE Transactions on Biomedical Engineering},
  volume       = {71},
  number       = {6},
  pages        = {1894--1905},
  year         = {2024},
  publisher    = {IEEE}
}

@article{fraiwan2021dataset,
  title        = {A Dataset of Lung Sounds Recorded from the Chest Wall Using an
                  Electronic Stethoscope},
  author       = {Fraiwan, Luay and Hassanin, Omnia and Fraiwan, Mohammad and
                  Khassawneh, Basheer and Ibnian, Ali M. and Alkhodari, Mohanad},
  journal      = {Data in Brief},
  volume       = {35},
  pages        = {106913},
  year         = {2021},
  publisher    = {Elsevier},
  doi          = {10.1016/j.dib.2021.106913}
}

@article{kandasamy2011monitoring,
  title        = {Monitoring and Analysis of Lung Sounds Remotely},
  author       = {Kandasamy, Thirumaran and Ramachandran, Krishnamurthy and
                  Lourdusamy, Paulose},
  journal      = {International Journal of Telemedicine and Applications},
  volume       = {2011},
  year         = {2011},
  publisher    = {Hindawi},
  doi          = {10.1155/2011/340861}
}

@article{zhang2022sprsound,
  title        = {{SPRSound}: Open-source {SJTU} Paediatric Respiratory Sound Database},
  author       = {Zhang, Qing and Zhang, Jing and Yuan, Jiajun and Huang, Huajie and
                  Zhang, Yuhang and Zhang, Baoqin and Lv, Gaomei and Lin, Shuzhu and
                  Wang, Na and Liu, Xin and others},
  journal      = {IEEE Transactions on Biomedical Circuits and Systems},
  volume       = {16},
  number       = {5},
  pages        = {867--881},
  year         = {2022},
  publisher    = {IEEE}
}

@inproceedings{sharma2020coswara,
  title        = {{Coswara -- A Database of Breathing, Cough, and Voice Sounds for
                  COVID-19 Diagnosis}},
  author       = {Sharma, Neeraj Kumar and Krishnan, Prashant and Kumar, Rohit and
                  Ramoji, Shreyas and Chetupalli, Srikanth Raj and Nirmala, R. and
                  Ghosh, Prasanta Kumar and Ganapathy, Sriram},
  booktitle    = {Proc. Interspeech 2020},
  pages        = {4811--4815},
  year         = {2020},
  doi          = {10.21437/Interspeech.2020-2768}
}

@inproceedings{orlandic2023multimodal,
  title        = {A Multimodal Dataset for Automatic {Edge-AI} Cough Detection},
  author       = {Orlandic, Lara and Thevenot, J{\'e}r{\^o}me and Teijeiro, Tomas and
                  Atienza, David},
  booktitle    = {2023 45th Annual International Conference of the IEEE Engineering
                  in Medicine \& Biology Society (EMBC)},
  pages        = {1--7},
  year         = {2023},
  publisher    = {IEEE},
  doi          = {10.1109/EMBC40787.2023.10340413}
}

@article{baur2024hear,
  title        = {{HeAR} -- Health Acoustic Representations},
  author       = {Baur, Sebastien and Nabulsi, Zaid and Weng, Wei-Hung and
                  Garrison, Jake and Blankemeier, Louis and Fishman, Sam and
                  Chen, Christina and Kakarmath, Sujay S. and Maimbolwa, Minyoi M. and
                  Sanjase, Nsala and others},
  journal      = {arXiv preprint arXiv:2403.02522},
  year         = {2024}
}

@article{zhang2024opera,
  title        = {Towards Open Respiratory Acoustic Foundation Models: Pretraining
                  and Benchmarking},
  author       = {Zhang, Yuwei and Xia, Tong and Han, Jing and Wu, Yunpeng and
                  Rizos, Georgios and Liu, Yang and Mosuily, Mohamed and
                  Chauhan, Jagmohan and Mascolo, Cecilia},
  booktitle    = {Advances in Neural Information Processing Systems (NeurIPS)},
  year         = {2024}
}

@article{zhang2024respllm,
  title        = {{RespLLM}: Unifying Audio and Text with Multimodal {LLMs} for
                  Generalized Respiratory Health Prediction},
  author       = {Zhang, Yuwei and Xia, Tong and Saeed, Aaqib and Mascolo, Cecilia},
  journal      = {arXiv preprint arXiv:2410.05361},
  year         = {2024}
}

@article{anonymous2025pcg,
  title        = {Exploring Fine-Tuned Audio-{LLM} on Heart Murmur Features},
  author       = {Anonymous},
  journal      = {arXiv preprint arXiv:2501.13884},
  year         = {2025}
}

@article{sellergren2025medgemma,
  title        = {{MedGemma} Technical Report},
  author       = {Sellergren, Andrew and Kazemzadeh, Sahar and Jaroensri, Tiam and
                  Kiraly, Atilla and Traverse, Madeleine and Kohlberger, Timo and
                  Xu, Shawn and Jamil, Fayaz and Hughes, C{\'\i}an and Lau, Charles
                  and others},
  journal      = {arXiv preprint arXiv:2507.05201},
  year         = {2025}
}

@inproceedings{radford2021clip,
  title        = {Learning Transferable Visual Models from Natural Language Supervision},
  author       = {Radford, Alec and Kim, Jong Wook and Hallacy, Chris and Ramesh, Aditya
                  and Goh, Gabriel and Agarwal, Sandhini and Sastry, Girish and
                  Askell, Amanda and Mishkin, Pamela and Clark, Jack and Krueger, Gretchen
                  and Sutskever, Ilya},
  booktitle    = {Proceedings of the 38th International Conference on Machine Learning
                  (ICML)},
  volume       = {139},
  pages        = {8748--8763},
  year         = {2021},
  publisher    = {PMLR}
}
\bibliographystyle{icml2026}

\end{document}